\documentstyle[preprint,aps,epsf]{revtex}
\tighten
\begin{document}
\draft
\title{Magnetic hysteresis in Ising-like dipole-dipole model}
\author{Gy\"orgy Szab\'o and Gy\"orgy K\'ad\'ar}
\address
{Research Institute for Technical Physics and Materials Science \\
POB 49, H-1525 Budapest, Hungary}
\date{\today}
\maketitle

\begin{abstract}
Using zero temperature Monte Carlo simulations we have studied the
magnetic hysteresis in a three-dimensional Ising model with nearest
neighbor exchange and dipolar interaction. The average magnetization
of spins located inside a sphere on a cubic lattice is determined 
as a function of magnetic field varied periodically. The simulations
have justified the appearance of hysteresis and allowed us to have a
deeper insight into the series of metastable states developed during
this process.
\end{abstract}

\section{INTRODUCTION}
\label{sec:intro}

The description of hysteresis---predominantly for magnetization
curves---has been the aim of numerous papers for more than a century now.
The early phenomenological model of Lord Rayleigh was the fundamental idea
of the well known Preisach-model of ferromagnetic hysteresis, which has
been further developed and widely discussed together with other
descriptive phenomenological models in a long series of works \cite{preis}
appearing up till the present days. The physical explanation for the lag
of the magnetization component behind the external magnetic field
varying along a given line was elaborated by Stoner and Wohlfarth \cite{SW}
in a simple micromagnetic model for the case of a single-domain particle
of uniform magnetization. With the advance of the technics of
statistical physics recently great interest seems to be oriented towards
the investigation of hysteretic phenomena in realistic multi-particle
and/or multi-domain systems. The aim after all would be narrowing the
gap between the phenomenological "top-down" and the physical "bottom-up"
approaches to the description of hysteresis in general.

In this paper Monte Carlo simulation of a multi-particle Ising system
of point-like elementary magnetic dipoles will be presented. The dipoles
contained inside a sphere are arranged in a three-dimensional simple
cubic lattice, they are parallel or antiparallel to the $z$-axis of the
Cartesian system and interact by the nearest neighbour exchange as well
as the long-range dipole-dipole couplings.

It will be shown that this model exhibits magnetic hysteresis if the
external magnetic field is varied periodically. The present approach
allows us to visualize and analyse the evolution of spin configurations.

\section{THE MODEL}
\label{sec:model}

We consider a three-dimensional Ising model with spins located on the points 
of a simple cubic lattice (${\bf r}=(x,y,z); \ x, y,$ and $z$ are integers)
within a sphere of radius $R$ ($r^2=x^2+y^2+z^2<R^2$). Dimensionless
quantities will be used throughout the paper. Namely, the length will be
measured in terms of lattice constant $a$, the dipole moments in terms of
the unique dipole moment $\mu$, the external magnetic field $h$ will be
expressed in units of $\mu/a^3$ and the energy per spins in terms of
$\mu^2/a^3$. In this case the Hamiltonian is given by
\begin{equation}
H=- J \sum_{\langle {\bf r, r^{\prime}} \rangle} \sigma({\bf r})
\sigma({\bf r^{\prime}})- h \sum_{\bf r} \sigma ({\bf r})
+{1 \over 2}\sum_{\bf r, r^{\prime} \atop r\ne r^{\prime}}
V({\bf r-r^{\prime}}) \sigma({\bf r}) \sigma({\bf r^{\prime}})
\label{eq:Hamil}
\end{equation}
where $\sigma ({\bf r})=1$ and -1 for up and down spins. In the first term
the sum is over the nearest-neighbor pairs and $J$ denotes the strength of
the usual exchange interaction measured in the above mentioned energy unit.
The second term describes the effect of the external magnetic field $h$.
The dipole-dipole interaction between two spins having only z component
is defined as
\begin{equation}
V({\bf r}) = {r^2-3z^2 \over r^5} \ \ .
\label{eq:Vdip}
\end{equation}

Notice that this dipole-dipole interaction provides ferromagnetic coupling
along the $z$ axis and the coupling becomes antiferromagnetic in the $x-y$
plane. Furthermore, the average value of the field of a given dipole over
the points located at a prescribed distance vanishes due to
the cubic symmetry. Conversely, up (or down) spins on a ``spherical shell''
produces zero magnetic field at the central lattice point.
This is the reason why the resultant magnetic field is zero at the center
of a spherical sample if all the spins point to up (or
down).\cite{sauer,kittel,lutt,cohen} At a given site ${\bf r}$ the magnetic
field produced by the remaining dipoles is given as
\begin{equation}
h_d({\bf r})= - \sum_{\bf r^{\prime} \atop r^{\prime}\ne r}
V({\bf r-r^{\prime}}) \sigma({\bf r^{\prime}}) \ .
\label{eq:h}
\end{equation}
Our analysis is restricted to spherical systems because here the early
theories predicts zero field inside the ferromagnetic sphere.
\cite{sauer,kittel,lutt,cohen} More precisely, Cohen and Keffer
\cite{cohen} have shown that the contribution of point-like dipoles to
the local field may differ from zero in the close vicinity of the surface.
We have numerically studied the local field because this phenomenon plays
crucial role in the magnetic hysteresis as will be described in the
next section.

\begin{figure}
\centerline{\epsfxsize=10cm
            \epsfbox{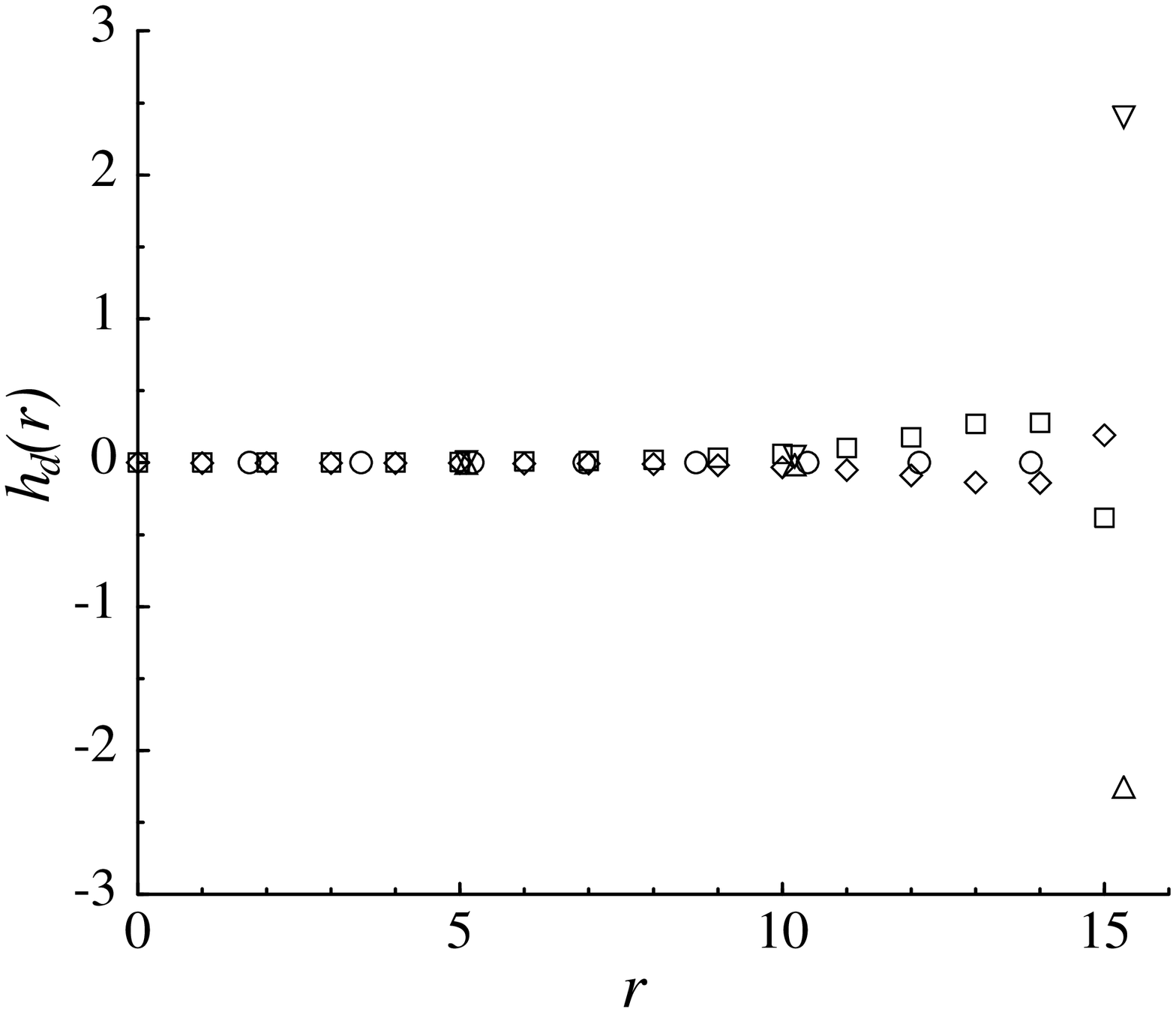}
            \vspace*{1mm}   }
\caption{Radial dependence of the local magnetic field if $m=1$ and
$R=15.33$ for choosing different directions: (1,0,0) diamonds, (0,0,1)
open squares, (1,1,1) open circles, (5,0,1) $\bigtriangleup$ and
(1,0,5) $\bigtriangledown$.}
\label{fig:hdr}
\end{figure}

In the ferromagnetic state the average $h_d({\bf r})$ (energy per sites)
due to dipolar interactions is zero \cite{lutt}
independently of the system size. In this case the local field satisfies
the conditions of reflection ($h_d(x,y,z)=h_d(\pm x,\pm y, \pm z)$) and
rotation ($h_d(x,y,z)=h_d(-y,x,z)$) symmetries. The numerical results
(see Fig.~\ref{fig:hdr}) demonstrate clearly that inside our sphere
$h_d({\bf r}) \approx 0$ in agreement with the predictions of analytical
calculations \cite{sauer,kittel,lutt,cohen}. At the same time large variations
are indicated on the outer shells. For $\sigma({\bf r})=+1$ the highest
values of $h_d$ appear along the periphery of the top and bottom layers.
Along the (1,1,1) directions the local field vanishes (see the open circles
in Fig.~\ref{fig:hdr}). The lowest field values are found at the 8 sites
symmetrically equivalent to (15,0,3) if $R=15.33$. 

The probability distribution of the lattice points as a function of $h_d$
exhibits a sharp peak around $h_d=0$, and its maximum value increases
with $R$. In a cube-shaped sample such calculation yields a significantly
different (wide) probability distribution which causes drastic changes
in the hysteresis too.

The dipolar energy of ordered spin configurations was determined previously
in infinite systems of cubic symmetry. Using the Ewald
method Luttinger and Tisza \cite{lutt} evaluated the energy per sites
for all the (periodic) ordered structures characterized by a spin
configuration within the $2\times 2\times 2$ unit cell. In the energetically
favoured states the up (or down) spins form vertical columns as
expected. For $h=0$ and $J=0$ the spin configuration of lowest energy is
a twofold degenerated chess-board-like antiferromagnetic arrangement in
the $x-y$ plane of ferromagnetic columns along the $z$ direction,
that is $\sigma ({\bf r})=1$ (-1) if $x+y$ is odd (even). In this columnar
antiferromagnetic (CAF) structure the energy per sites is given as
\cite{lutt}
\begin{equation}
E_{CAF}= -2.676+J \ .
\label{eq:ECAF}
\end{equation}
Furthermore, for the fourfold degenerated layered antiferromagnetic (LAF) spin
configuration, where $\sigma ({\bf r})=1$ (-1) if $x$ (or $y$) is odd
(even), the energy per sites is
\begin{equation}
E_{LAF}= -2.422-J \ .
\label{eq:ELAF}
\end{equation}
It is obvious from Eqs.~(\ref{eq:ECAF}) and (\ref{eq:ELAF}) that the CAF
state is preferred to LAF if the ferromagnetic coupling $J<0.127$.
In our spherical model numerical calculations are performed to check the
finite size effects choosing $R=10.25$, 12.25 and 15.33. For both the above
mentioned spin configurations it is found that finite size corrections are
proportional to $1/R$. The numerical technique has allowed us to study
periodic antiferromagnetic structures different from those studied by
Luttinger and Tisza\cite{lutt}. These calculations indicate that double or
multilayer structures (similar to LAF) become stable for sufficiently strong
ferromagnetic coupling and the preferred layer width increases with $J$.
These indications, however, should be considered as preliminary results
because more systematic analyses are required to investigate the effect of
(anti)ferromagnetic coupling on the spin configurations in the ground state,
the size effects, {\it etc}. Henceforth, we will concentrate on the model
with $J=0$. In this case the slow cooling Monte Carlo technique has confirmed
that the twofold degenerated ground state is equivalent to the CAF spin
configuration in zero external field ($h=0$).

\section{SIMULATION OF HYSTERESIS}
\label{sec:sim}

\begin{figure}
\centerline{\epsfxsize=10cm
            \epsfbox{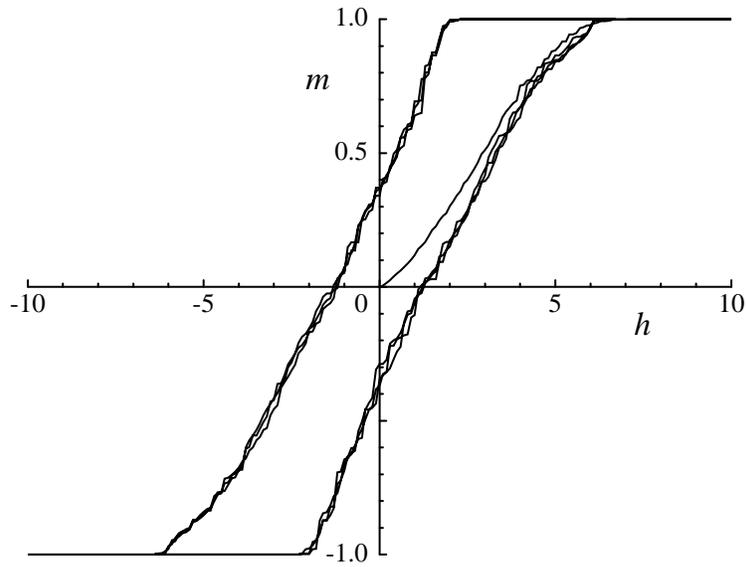}
            \vspace*{1mm}    }
\caption{Magnetic hysteresis for $J=0$ and $R=15.33$.}
\label{fig:hyst}
\end{figure}

A series of zero temperature Monte Carlo simulations has been performed
to study hyteresis phenomena in the model described above. In these
simulations the system is started from a random spin configuration.
For an elementary process a randomly chosen spin is flipped if
$(h_d({\bf r})+h)\sigma({\bf r})<0$, otherwise the spin value remains
unchanged. This process is repeated until all the spin signs become
equivalent to the sign of the local magnetic field. Then the external
magnetic field ($h$) is increased (decreased) by $\Delta h$ and repeating
the above mentioned spin flipp processes the system is allowed to relax
toward a new local energy minimum. In such a way the external field is varied
periodically with an amplitude of $10$. During this procedure we monitored
the magnetization defined as
\begin{equation}
m={1 \over N} \sum_{\bf r} \sigma({\bf r})
\label{eq:m}
\end{equation}
where $N$ is the number of spins inside the sphere. These simulations have
clearly indicated the appearance of the usual magnetic hysteresis. A
typical result obtained for $J=0$, $R=15.33$ and $|\Delta h|=0.1$ is shown
in Fig.~\ref{fig:hyst}. To check the size effect the simulations have been
carried out for different sizes ($R=10.25$ and 12.25). The comparisons have
indicated that the size effect is comparable to the uncertianty of data
observed between subsequent cycles (see Fig.~\ref{fig:hyst}). In other words,
the above mentioned system size containing $15\,203$ spins is sufficiently
large to study the hysteresis adequately in this model. Unfortunately, for
larger sizes the simulation becomes very time-consuming because the
computational time is proportional to $R^6$. The choice of
$|\Delta h|=0.1$ is also motivated by the minimization of run time. Further
decrease of $\Delta h$ does not modify the plotted curves significantly.

Figure \ref{fig:hyst} indicates the presence of avalanches when varying
the magnetic field. Similar phenomena were observed in experiments by
Barkhausen \cite{bark} and also in the random field Ising models
\cite{RFI}. Recently different approaches are used to study the avalanches
as well as its relation to the hysteresis.\cite{nara,bert} In the
present model the appearence of avalanches is a consequence of dipole-dipole
interaction.

\begin{figure}
\centerline{\epsfxsize=16cm
            \epsfbox{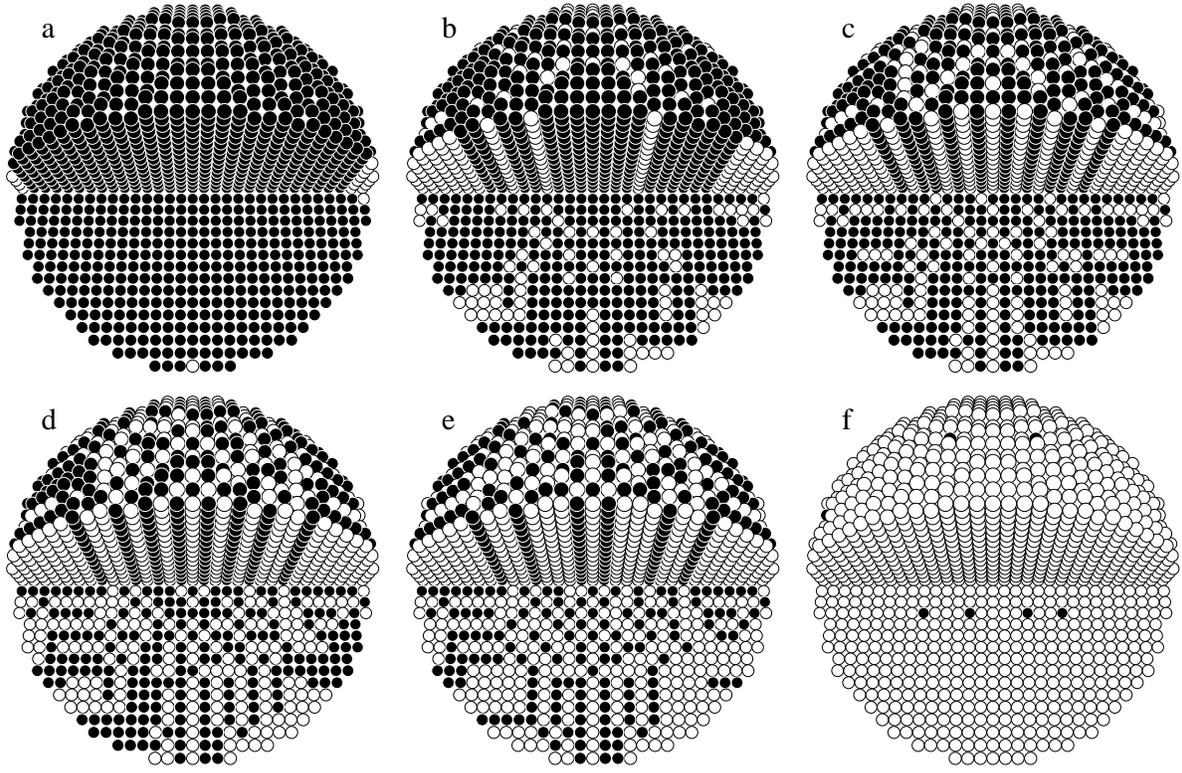}
            \vspace*{1mm}    }
\caption{Spin configurations when decreasing the magnetic field from
$h=10$ to 2 (a); 1.7 (b); 0 (c); -1 (d); -4 (e) and -6 (f). Black and
white spheres indicate up and down spins.}
\label{fig:cfgs}
\end{figure}

By varying the magnetic field several spins of the actual configuration
becomes preferred to flip into the opposite state. Reversing a randomly
chosen spin the local field $h_d({\bf r})$ will be modified in all sites
of the system. Due to the dipole-dipole interaction the neighboring spins
in the same column are favoured to change direction too. This
effect drives the avalanche of spin flips in a given column. During the
simulations the spin flips are observed in several columns simultaneously.
This process leads predominantly to such configurations where complete
spin columns have been reversed as demonstrated in Fig.~\ref{fig:cfgs}.

Figure \ref{fig:cfgs} shows six spin configurations appeared at different
magnetic fields during a half-cycle when decreasing $h$ from 10.
In order to visualize a more complete 3D picture on the spin configurations
parts of the horizontal ($z=0$) and vertical $(x=0)$ cross sections
are displayed by removing a quarter of spins from the sphere. In this
figure the small black spheres with white border and the white spheres with
black border indicate up and down spins.

Decreasing the magnetic field the initial saturated state ($m=1$) remains
unchanged until $h=2.253$ whose value depends slightly on $R$.
For the given size the spins which can be the first to flip are positioned
at one of the 8 sites equivalent to $(15,0,3)$. Consequently, the 
spins in the corresponding four columns can flip simultaneously,
because the short columns are too far to affect each other significantly.
The result is well recognizable for a possible subsequent configuration
plotted in picture $a$. Notice that here the mentioned symmetries are
no longer valid due to a branching process described as follows. 
Immediately after the decrease of $h$ some of the spins become reversible.
One of them is chosen randomly to flip. This elementary process fastened
by the above mentioned columnar spin flip avalanche can actually prevent
the flips at the ``rival spins''. Thus the order of consecutive steps during
the formation of the subsequent columnar structures (see Fig.~\ref{fig:cfgs})
is occasional. This phenomenon causes the hysteresis curves to be
unreproducable for finite sizes as demonstrated in Fig.~\ref{fig:hyst}.

Configuration $c$ in Fig.~\ref{fig:cfgs} represents a typical state
with a remanent magnetization when reaching $h=0$. The magnetization
vanishes at $h \simeq -1$. Here, the appearence of domains with CAF
and LAF structures are striking (see configuration $d$). The CAF state
dominates the vicinity of the $z$ axis. The positions of the LAF domains are
initialized by the first columnar spin flips after saturation. These domains
remains recognizable in a wide range of the magnetization process as indicated 
by the configuration $e$ obtained for $h=-2$. Further decrease of $h$ will
destroy these ordered regions leaving the spins unchanged only in a
few columns positioned randomly (configuration $f$ for $h=-6$). Notice
that this configuration differs significantly from the initial ones
(compare to $a$). Finally all the spins point to downward if the magnetic
field becomes less than -6.2.

When studying a more complete series of spin configurations one can easily
recognize that the back spin flips appears rarely upon a half-cycle.
According to our numerical investigations the number of back spin
flips is less than 1\% of the total number of spins. During the simulations
we have also recorded the number of non-uniform columns which is found
to be zero in the initial and final stages of demagnetizing processes. This
number can occasionally differ from zero and its rate remains below 1\%.

The low number of back flips explains the phenomena observed when we have
articially prevented the free spin flips. In this case the above described
dynamics is modified by assuming a hysteretic behavior for the individual
spins. Namely, the spin flips may occur for the randomly chosen sites only
if the driving force exceeds a threshold value ($f_t \geq 0$). Evidently,
for $f_t=0$ this modification leaves the above
results unchanged. In agreement with the expectation the hysteresis loop
becomes wider for pozitive $f_t$. More precisely, the ``sides'' of the
loops are shifted outward without causing any observable changes in the
slopes. This means that the main features of the subsequent spin
configurations are similar to those described above (see Fig.~\ref{fig:cfgs})
and the columnar spin flips are delayed within the cycles. However, the
number of back flips (during the half-cycle) decreases with $f_t$.
For example, only a few back flips can be observed for $R=15.33$
and $f_t=1$ and this elementary process vanishes practically if $f_t>2$.

\begin{figure}
\centerline{\epsfxsize=10cm
            \epsfbox{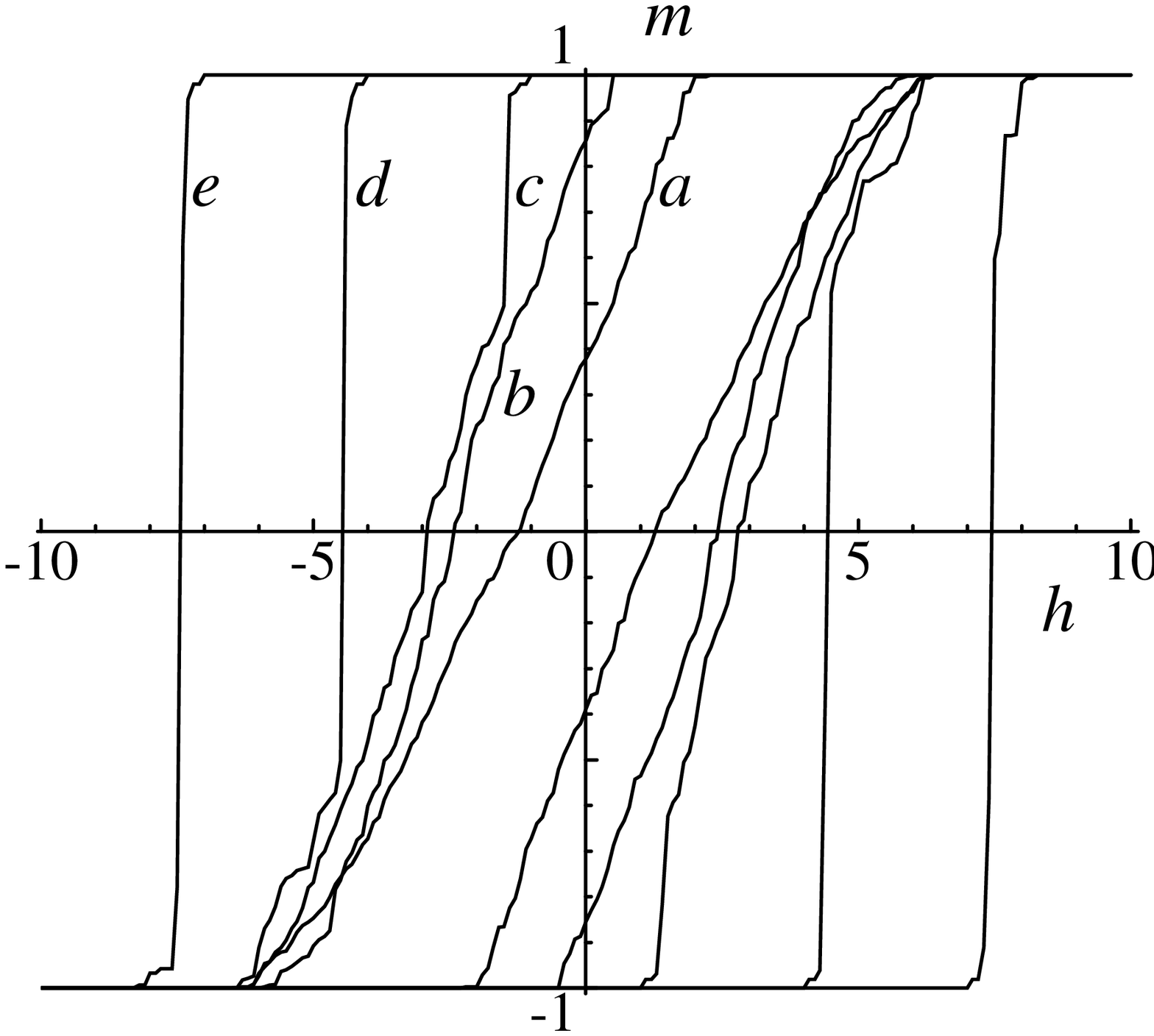}
            \vspace*{1mm}   }
\caption{Magnetic hysteresis curves for different exhange constants:
$J=0$ (a), 0.5 (b), 1 (c), 2 (d) and 3 (e) for  $R=15.33$. The curves are
averaged over 10 cycles.}
\label{fig:hyst-J}
\end{figure}

Finally we have investigated the effect of ferromagnetic coupling ($J>0$)
on the hysteresis. For this purpose the hysteresis curves are recorded
for different $J$ values as shown in Fig.~\ref{fig:hyst-J}. In this case
the simulations are started from the ferromagnetic state ($h=10$). We have
observed that the extension of avalanches (as well as the uncertainty
of the magnetization process for the subsequent cycles) increases with $J$.
Figure~4 demonstrates that these phenomena are accompaned
with the increase of average slope along the ``side'' of loops and the
broadening of the hysteresis loop.
These observations are related to formation of larger and larger
ferromagnetic domains as well as the more and more massive avalanches
when increasing the ferromagnetic coupling.

One can observe that the initial steps of the demagnetization process
is slightly modified by the nearest-neighbor couplings because the first
spin flips appear on the surface at low latitudes where $h_d({\bf r})$ has
low (negative) values. This is the region where the ferromagnetic force
is weak because of the low number (three or four) of neighboring spins,
therefore it is not able to prevent the spin flips driven by
the local field $h+h_d({\bf r})$. Further systematic analysis is required
to study what happens when the typical ferromagnetic domain sizes become
larger than $R$.

\section{CONCLUSIONS}
\label{sec:conc}

Numerical simulations have been performed to investigate the hysteresis
phenomena in a three-dimensional Ising model involving both the nearest
neighbor exchange and the long range dipolar interactions.
For simplicity the spins are located on a cubic lattice within a sphere.
This shape provides zero local field $h_d({\bf r})$ inside the
sphere for a ferromagnetic state. The large deviations of $h_d({\bf r})$
at the surface are found to play crucial role at the beginning of
demagnetization processes started from a saturated state.
Using zero temperature Monte Carlo simulations the magnetization process
is investigated under an alternating external magnetic field with an
amplitude providing complete saturations.

During this process we have observed avalanches and branching points
leading to a large variation of paths along which the system can evolve.
The avalanches consist of spin flips constrained into a few columns.
As a result, after having varied the magnetic field the new stationary
(metastable) states built up dominantly from columns with uniform spins.
This feature decreases drastically the number of possible metastable state
(characterizing the hysteretic behaviour) in comparison to the total
number of configurations. Our simulations have justified that the number of
back flips is negligible within the half-cycles of magnetization process.

We think that the present model seems to be a good candidate for exploring
relationship between a microscopic description and the phenomenological
(e.g. Preisach model) approaches of hysteresis. For this purpose, however, 
we need more detailed information about the internal hysteresis loops,
the effects of shape and exchange constant, the local field distribution,
etc. Fortunately, the model simulations give us a wide scale of opportunity
to have a deeper understanding.

\acknowledgements

This research was supported by the Hungarian National Research Fund
(OTKA) under Grant No. T-23555.

\begin{references}

\bibitem{preis}Lord Rayleigh, Phil. Mag., {\bf 23}, 225 (1887); F. Preisach,
Z. Physik, {\bf 94}, 277 (1935); A. Visintin, {\it Differential Models of
Hysteresis}, (Springer-Verlag, Berlin Heidelberg, 1994).

\bibitem{SW}E. C. Stoner  and  E. P. Wohlfarth,  Phil.\ Trans.\ Roy.\ Soc.\
(London), {\bf A-240}, 599 (1948).

\bibitem{sauer}J. A. Sauer, Phys.\ Rev.\ {\bf 57}, 142 (1940).

\bibitem{kittel}C. Kittel, {\it Introduction to Solid State Physics},
5th edition (Wiley, New York, 1976).

\bibitem{lutt}J. M. Luttinger and L. Tisza, Phys.\ Rev.\  {\bf 70},
954 (1946); {\bf 72}, 257 (1947).

\bibitem{cohen}M. H. Cohen and F. Keffer, Phys.\ Rev.\ {\bf 99},
1128 (1955).

\bibitem{bark}Z. Barkhausen, Z. Phys.\ {\bf 20}, 401 (1919).

\bibitem{RFI}J. P. Sethna, K. Dahmen, S. Kartha, J. A. Krumhansl,
B. W. Roberts, J. D. Shore, Phys.\ Rev.\ Lett.\ {\bf 70}, 3347 (1993);
K. Dahmen, S. Kartha, J. A. Krumhansl, B. W. Roberts, J. P. Sethna,
and J. D. Shore, J. Appl.\ Phys.\ {\bf 75}, 5946 (1994); K. Dahmen
and J. P. Sethna, Phys.\ Rev.\ Lett.\ {\bf 71}, 3222 (1993);
O. Perkovi\'c, K. Dahmen, and J. P. Sethna, Phys.\ Rev.\ Lett.\
{\bf 75}, 4528 (1995).

\bibitem{nara}O. Narayan, Phys.\ Rev.\ Lett.\ {\bf 77}, 3855 (1996).

\bibitem{bert}G. Bertotti, Phys.\ Rev.\ Lett.\ {\bf 76}, 1739 (1996).


\end{references}
\end{document}